\documentclass[12pt]{article}
\usepackage{bbm}
\usepackage{epsfig}
\usepackage{array}
\usepackage{float} 
\usepackage{amsmath} 
\usepackage{amssymb}
\usepackage{amsfonts}

\usepackage{a4}
\usepackage{epsfig}

\usepackage{a4wide}
\usepackage{wasysym}

\def\ra{\rightarrow}
\def\be{\begin{equation}}
\def\ee{\end{equation}}
\def\gs{\mathrel{
   \rlap{\raise 0.511ex \hbox{$>$}}{\lower 0.511ex \hbox{$\sim$}}}}
\def\ls{\mathrel{
   \rlap{\raise 0.511ex \hbox{$<$}}{\lower 0.511ex \hbox{$\sim$}}}}

\newcommand{\ba}{\begin{array}{c}}
\newcommand{\baz}{\begin{array}{cc}}
\newcommand{\bad}{\begin{array}{ccc}}
\newcommand{\bea}{\begin{equation} \begin{array}{c}}
\newcommand{\eea}{ \end{array} \end{equation}}
\newcommand{\ea}{\end{array}}
\newcommand{\D}{\displaystyle}
\newcommand{\dms}{\mbox{$\Delta m^2_{\odot}$}}
\newcommand{\dma}{\mbox{$\Delta m^2_{\rm A}$}}

\begin{document}

\title{\vspace{-2cm}
\hfill {\small hep--ph/0703171} 
\vskip 0.6cm
\bf Large and Almost Maximal Neutrino Mixing
within the Type II See-Saw Mechanism}
\author{
Manfred Lindner\thanks{email: \tt lindner@mpi-hd.mpg.de}~
\mbox{ }~~and~~
Werner Rodejohann\thanks{email: \tt werner.rodejohann@mpi-hd.mpg.de} 
\\\\
{\normalsize \it Max--Planck--Institut f\"ur Kernphysik,}\\
{\normalsize \it Saupferecheckweg 1, D--69117 Heidelberg, Germany}\\ \\
}
\date{}
\maketitle
\thispagestyle{empty}
\vspace{-0.8cm}

\begin{abstract}
\vspace*{0.2cm}
\noindent 
Within the type II see-saw mechanism the light neutrino mass matrix is 
given by a sum of a direct (or triplet) mass term and the conventional 
(type I) see-saw term. Both versions of the see-saw mechanism explain
naturally small neutrino masses, but the type~II scenario offers 
interesting additional possibilities to explain large or almost maximal
or vanishing mixings which are discussed in this paper.
We first introduce ``type II enhancement'' of neutrino mixing, where 
moderate cancellations between the two terms can lead 
to large neutrino mixing even if all individual mass matrices and terms 
generate small mixing. However, nearly maximal or vanishing mixings are
not naturally explained in this way, unless there is a certain initial 
structure (symmetry) which enforces certain elements of the matrices 
to be identical or related in a special way. 
We therefore assume that the leading structure of the neutrino 
mass matrix is the triplet term and 
corresponds to zero $U_{e3}$ and maximal $\theta_{23}$. 
Small but necessary corrections are 
generated by the conventional see-saw term. 
Then we assume that one of the two terms corresponds to an extreme mixing 
scenario, such as bimaximal or tri-bimaximal mixing. Deviations from this 
scheme are introduced by the second term. 
One can mimic Quark-Lepton Complementarity in this way. 
Finally, we note that the neutrino mass matrix for tri-bimaximal mixing 
can be -- depending on the mass hierarchy -- 
written as a sum of two terms with simple structure. 
Their origin could be the two terms of type II see-saw.

\end{abstract}

 \newpage 

\section{\label{sec:intro}Introduction}

The smallness of neutrino masses arises naturally in the conventional 
or type~I see-saw mechanism \cite{I}, with a low energy neutrino mass 
matrix of the form 
\be \label{eq:I}
m_\nu^I = - m_D^T \, M_R^{-1} \, m_D ~. 
\ee
Here $m_D$ is a Dirac mass matrix usually related to the known 
fermion masses or the weak scale $v \simeq 174$ GeV, 
and $M_R$ is a Majorana mass matrix of Standard Model singlet 
neutrinos with a mass scale $M$ as large as the GUT scale. 
Hence, neutrino masses are naturally of order $v^2/M \sim 0.01$ eV, 
corresponding nicely to the square root of the mass squared 
difference of atmospheric neutrinos. However, the other equally 
astonishing aspect of neutrino physics, namely the presence of 
large mixing angles, is a priori not explained by the type~I 
see-saw mechanism. It is possible to generate large mixings by 
specific forms of the low energy mass matrix, and many models 
have been proposed \cite{review} in order to explain the form of 
$m_\nu$ from the structure of $m_D$ or of $M_R$, or of both of them. 

In this article we want to discuss large mixings in the context of 
the type II see-saw mechanism \cite{II}, where the light neutrino 
mass matrix can be written as the conventional type~I see-saw term 
plus an additional (triplet) contribution: 
\be \label{eq:II}
m_\nu = m_\nu^{II} + m_\nu^I = 
m_L - m_D^T \, M_R^{-1} \, m_D ~. 
\ee 
Since $m_\nu$ is now a sum of two terms, there are interesting 
non-trivial possibilities, not present in the conventional see-saw 
mechanism, which can naturally be related to large or nearly maximal 
mixings. The first suggestive option is, that for some reason, 
both terms could be of comparable magnitude and (moderate) 
cancellation is connected to the interesting features of neutrino 
mixing. Alternatively, it could be the sum of the two comparable 
terms which is crucial. Most naturally, one term dominates, while 
the other term introduces only a small correction. 
The interplay of both terms has so far been analyzed only in a 
few papers, for instance within specific $SO(10)$ models \cite{SO10}, 
regarding the reconstruction of the mass matrices \cite{IIreco0,IIreco}, 
or in other scenarios \cite{others,Xing,ich2,anki}. 
Specifically, we focus our discussion in this paper in the context 
of the type~II see-saw on four aspects of large neutrino mixing:
\begin{itemize}
\item[(i)] we point out in Section \ref{sec:I} that even if all 
involved matrices, and even both terms in Eq.~(\ref{eq:II}), 
generate small mixing, a moderate cancellation can generate 
large mixings in $m_\nu$ (``type II enhancement''). 
This happens if the involved matrices 
have similar or even identical flavor structure, which distinguishes 
the scenario from the usual (type~I) see-saw enhancement of neutrino 
mixing. This mechanism produces typically sizable or large mixings,
but maximal or exactly vanishing mixings are not expected  
(as in basically all models for lepton mixing) unless 
in addition certain elements of the matrices are related;  
\item[(ii)] in order to explain naturally almost maximal or almost
vanishing mixings, we propose in Section \ref{sec:II} that one of 
the two terms in $m_L - m_D^T \, M_R^{-1} \, m_D$ is dominant and
that it corresponds to $U_{e3}=0$ and to maximal $\theta_{23}$. 
The second term would then be responsible for small or tiny corrections; 
\item[(iii)] in Section \ref{sec:III} we assume that the triplet term 
in the type~II see-saw formula corresponds to a specific mixing scheme, 
e.g., bimaximal or tri-bimaximal mixing. A subleading conventional 
term then introduces a perturbation to this mixing scheme, thereby 
explaining deviations from bimaximal or tri-bimaximal mixing. It is 
also possible to mimic Quark-Lepton Complementarity in this way; 
\item[(iv)] in Section \ref{sec:tbm} we finally take advantage of the 
fact that the neutrino mass matrix 
for tri-bimaximal mixing can almost always be 
written as a sum of two terms with simple structure. 
Their origin could be the two terms of type II see-saw. 
\end{itemize}
To the best of our knowledge, the main points we make here have not been 
emphasized in the literature before. 
We will not construct explicit models for the issues given here, 
or conduct detailed numerical or analytical studies, but rather 
limit ourself to give instructive examples for each of these cases.  
We hope this will point the way to interesting model building possibilities and 
bring some attention to the various unexplored features of the type II 
see-saw. 
Before discussing the issues mentioned above, we will start by 
shortly summarizing the framework of the present study 
in the next Section.

\section{\label{sec:frame}Framework}
\subsection{\label{sec:mnugenn}Neutrino Mixing and the Mass Matrix}
Let us shortly summarize the neutrino observables and our current 
knowledge about them. 
With the usual parametrization of the lepton mixing matrix 
(we neglect the phases in this work), 
\bea \label{eq:Upara}
U = \left( \bad 
c_{12} c_{13} & s_{12} c_{13} & s_{13}  \\[0.2cm] 
-s_{12} c_{23} - c_{12} s_{23} s_{13}  
& c_{12} c_{23} - s_{12} s_{23} s_{13}  
& s_{23} c_{13}  \\[0.2cm] 
s_{12} s_{23} - c_{12} c_{23} s_{13}  & 
- c_{12} s_{23} - s_{12} c_{23} s_{13}  
& c_{23} c_{13}  \\ 
               \ea   \right)~, 
\eea
where $c_{ij} = \cos\theta_{ij}$, 
$s_{ij} = \sin\theta_{ij}$, 
we have as best-fit points \cite{thomas} 
$U_{e3}=0$, $\theta_{23}=\pi/4$ and $\sin^2 \theta_{12} \simeq 0.3$. 
The zeroth order form of the mass matrix 
$m_\nu = U^\ast \, m_\nu^{\rm diag} \, U^\dagger$ can then be 
given as 
\be \label{eq:leading}
m_\nu = \sqrt{\frac{\dma}{4}} \, 
\left( 
\bad
0 & 0 & 0 \\[0.2cm]
\cdot & 1 & 1 \\[0.2cm]
\cdot & \cdot & 1 
\ea
\right) \mbox{ or } m_\nu = \sqrt{\frac{\dma}{2}} \, 
\left( 
\bad
0 & 1 & 1 \\[0.2cm]
\cdot & 0 & 0 \\[0.2cm]
\cdot & \cdot & 0 
\ea
\right)~,
\ee
when neutrinos obey a normal ($m_3^2 \gg m_{2,1}^2$) or inverted 
($m_2^2 \simeq m_1^2 \gg m_3^2$ with $m_1$ and $m_2$ having 
opposite $CP$ parities) hierarchy, respectively. 
Order one coefficients are not explicitly given here. 
The matrices in 
Eq.~(\ref{eq:leading}) can for instance be obtained by asking for the 
conservation of the flavor charge $L_e$ \cite{le} or 
$L_e - L_\mu - L_\tau$ \cite{lelmlt}, respectively. 
Approximately, the dominating 23 block of the mass matrix can also be 
generated by sequential dominance of the right-handed neutrinos 
in type I see-saw scenarios \cite{sedom}.
If neutrinos are quasi-degenerate, $m_3 \simeq m_2 \simeq m_1 \equiv m_0$, 
there are also ways to 
explain this by simple symmetries. 
For instance, models based on $SO(3)$ usually lead to a 
mass matrix proportional to the unit matrix \cite{so3}, 
i.e., the three neutrinos all have the same $CP$ parities. 
This is a rather unstable situation in 
what regards radiative corrections. 
Another possibility, along the lines of 
$L_e$ and $L_e - L_\mu - L_\tau$, is 
\be \label{eq:lmlt}
m_\nu = m_0 \, 
\left( 
\bad
1 & 0 & 0 \\[0.2cm]
\cdot & 0 & 1 \\[0.2cm]
\cdot & \cdot & 0 
\ea
\right)~,
\ee
which corresponds to the conservation of $L_\mu - L_\tau$ \cite{lmlt0,lmlt}. 
Apparently, some or all of the zero elements of these simple matrices 
in Eqs.~(\ref{eq:leading}, \ref{eq:lmlt}) have to be filled with 
small entries. Alternatively, the flavor symmetries $L_e$, 
$L_e - L_\mu - L_\tau$ or $L_\mu - L_\tau$ have to be broken softly. 
One of the points we wish to make in this paper is that 
the type II see-saw mechanism with its two terms generating $m_\nu$ 
is a natural candidate to introduce the breaking parameters.

\subsection{\label{sec:IIgen}Origin of Type II See-Saw}

The low energy neutrino mass matrix resulting from the type II see-saw is 
$m_\nu = m_\nu^{II} + m_\nu^I = m_L - m_D^T \, M_R^{-1} \, m_D$. 
The relevant Lagrangian is
\be \label{eq:L}
{\cal L} = \frac 12 \, \overline{N_{R i}} \, (M_R)_{ij} \, N_{Rj}^c + 
\frac 12 \, \overline{L_\alpha^c} \, 
f_{\alpha \beta} \, i \tau_2 \, \Delta_L \, L_\beta 
+ \frac 1v \, 
\overline{N_{R i}} \, (m_D)_{i \alpha} \, L_\alpha \, \Phi^\dagger ~, 
\ee
where $N_{R i}$ are the right-handed Majorana neutrinos and 
$L_\alpha = (\nu_\alpha, \alpha)_L^T$ is the lepton doublet 
with $\alpha = e, \mu, \tau$. 
There is also a Dirac mass matrix $m_D$  
governing the coupling of the Higgs doublet $\Phi$ with the $N_{R i}$. 
The matrix $m_\nu^{II}$ stems from the second term in Eq.~(\ref{eq:L}) 
and requires a $SU(2)_L$ triplet, which can be written as 
\be
\Delta_L = 
\left( 
\bad 
\frac{1}{\sqrt{2}} \, \Delta^+ & \Delta^{++} \\[0.2cm]
\Delta^0 & -\frac{1}{\sqrt{2}} \, \Delta^+
\ea
\right)~.
\ee
The neutral component develops a vacuum expectation value $v_L$, which 
together with the symmetric Yukawa coupling matrix $f_{\alpha \beta}$ 
gives a contribution $m_\nu^{II} = v_L \, f$ to the low energy 
neutrino mass matrix. The value of the $\rho$ parameter and in particular 
the small neutrino masses imply that $v_L \ll v$. 
A popular scenario in which the type II see-saw can be 
realized is based on the left-right (LR) symmetric gauge group 
$SU(2)_L \times SU(2)_R \times U(1)_{B - L}$. 
The LR gauge group is a subgroup of the Pati-Salam group and it can also 
be obtained from $SO(10)$. 
Gauge symmetry implies the existence of $V - A$ and $V + A$ interactions. 
Moreover, gauge symmetry demands the presence of a 
$SU(2)_R$ Higgs triplet $\Delta_R$. 
By developing a vacuum expectation value $v_R$, 
$SU(2)_L \times SU(2)_R \times U(1)_{B - L}$ is broken down 
to the Standard Model. The mass matrix of the right-handed 
neutrinos $M_R = v_R \, g$ is also generated, 
where $g$ is a symmetric Yukawa coupling 
matrix\footnote{The Higgs doublet now becomes a bi-doublet, 
which however does not affect our discussion.}. 
An even more appealing and interesting scenario occurs 
when in addition to the LR gauge symmetry there is a 
discrete LR symmetry, in which case 
$m_L$ and $M_R$ have identical flavor structure and are 
proportional to each other: 
\be
m_L \equiv v_L \, f 
= \frac{v_L}{v_R}  \, M_R~\mbox{ with } v_L \, v_R = \gamma \, v^2~,   
\ee
where $\gamma$ is a model-dependent function of the underlying theory. 
The discrete LR symmetry of the form $f = g$ implies in addition that 
$m_D$ is symmetric. 
The Yukawa matrix $f$ defines the flavor structures of both 
$m_L$ and $M_R$. 
Using $v_L \, v_R = \gamma \, v^2~$ we have 
\be \label{eq:mnuLR}
m_\nu = v_L \, \left( f - m_D^T \, \frac{f^{-1}}{\gamma \, v^2} \, m_D 
\right)~.
\ee
It is apparent that the relative magnitude of the two terms in 
$m_\nu$ depends on $\gamma$. 
In order to have both terms in the type II see-saw formula to be of 
similar magnitude, and with assuming that at least one entry of $m_D$ is of 
order $v$, the value $\gamma = {\cal O}(1)$ suggests 
itself\footnote{Actually, the situation is slightly more 
complicated \cite{IIreco}, but this simplified discussion suffices 
to emphasize the main points.}. 
Moreover, if one entry of $m_D$ is of order $v$, dominance of the 
conventional see-saw term corresponds to $\gamma \ll 1$, 
whereas dominance of the triplet term corresponds to $\gamma \gg 1$. 
In the limit of $v_R \ra \infty$ the parameter $v_L$ and therefore the 
neutrino mass goes to zero. In addition, 
the theory becomes purely $V - A$. Hence, such theories relate the 
smallness of neutrino 
masses with the maximal parity violation of the weak interactions, a feature 
which makes them from an esthetical point of view very 
attractive. Actually, to have this connection a LR gauge symmetry suffices 
and no need for a discrete symmetry is present. 
In fact, for most of the issues to be discussed in the following, 
neither a LR gauge nor discrete symmetry are necessary. 
From the model building point of view it is however interesting to see 
where one could afford such a symmetric and esthetical framework. 
Moreover, gauge and discrete LR symmetry reduce the number of free 
parameters and simplify the analysis. 
Some cases to be presented will however not 
be possible when a discrete left-right symmetry is present.\\

The possibility that a term is added to the conventional 
see-saw term $m_D^T \, M_R^{-1} \, m_D$ 
is of course not exclusively reserved for a Higgs triplet.  
There can be 
$B - L$ breaking dimension five operators from various sources \cite{ss25}, 
including Planck scale effects, SUSY contributions, 
radiative models, etc. All of these possibilities have their theoretical 
justification, and in principle our considerations can apply to these 
contributions, too.

\section{\label{sec:I}From Small to Large Mixing via Type II See-Saw}
In this Section we remark that in the type II see-saw mechanism moderate 
cancellation can lead to the generation of large neutrino mixing. 
This mechanism, which we call ``type II enhancement'' 
of neutrino mixing, can work successfully even if $m_D$, $m_L$, $M_R$ 
and $m_D^T \, M_R^{-1} \, m_D$ 
correspond to small mixing. In fact, we will assume here that all 
individual matrices possess 
a ``hierarchical masses with small mixing'' form. 
This has its motivation in the large hierarchy of the charged lepton 
and quark masses, as well as the small quark mixing. 

Let us first recall the generation of large mixing from small 
mixing in case of the conventional see-saw mechanism. 

\subsection{The Situation in the Type I See-Saw}

Before experimental results made a paradigm change necessary, 
one expected that there is some form of quark-lepton symmetry 
which then implies that lepton mixing -- in analogy to quark mixing --
is described by small mixing angles. 
However, after the discovery of large lepton mixing it turned out that 
in principle one can 
generate large mixing in $m_\nu^I$ from small mixing in $m_D$ and $M_R$ 
by appropriate choice of the hierarchies in, and parameters of, 
the matrices $m_D$ and $M_R$ \cite{largeI}. 
For instance, in a 
simple 2-flavor framework the mass matrices could be
\be
m_D = v \, 
\left( 
\baz
\epsilon_D & a \, \epsilon_D \\[0.2cm]
b \, \epsilon_D & 1 
\ea
\right) \mbox{ and } 
M_R = M \, 
\left( 
\baz
\epsilon_M & 0 \\[0.2cm]
\cdot & 1 
\ea
\right)~,
\ee
with $\epsilon_{D,M} \ll 1$ and $a,b = {\cal O}(1)$. The individual 
mixing 
angles of these two matrices are small or even zero. 
The relevant parameter for the relative hierarchy between $m_D$ and 
$M_R$ is $\eta \equiv \epsilon_D^2 /\epsilon_M$. In case of 
$\eta \gg 1$, or $\epsilon_D^2 \gg \epsilon_M$, 
the mixing angle for $m_\nu = -m_D^T \, M_R^{-1} \, m_D$ 
is large: 
\be \label{eq:mnu22}
m_\nu = -\frac{v^2}{M}
\left( 
\baz 
\eta + b^2 \, \epsilon_D^2  & 
a \, \eta + b \, \epsilon_D \\[0.2cm]
\cdot & 1 + a^2 \, \eta 
\ea 
\right) \stackrel{\eta \gg 1}{\Longrightarrow} 
\tan 2 \theta \simeq \frac{2}{a - 1/a} = {\cal O}(1)~.
\ee
Note that the individual mixings of $m_D$ and $M_R$ are small 
-- in analogy to the quark sector -- but the mixing 
of $m_\nu$ is large. This ``see-saw enhancement''  
can be traced to $\epsilon_D^2 \gg \epsilon_M$, 
i.e., a stronger hierarchy in the Majorana sector \cite{largeI}. Naively, 
one might say that the hierarchy of $m_D$ is squared in the type 
I see-saw formula, so that the hierarchy in $M_R$ has to be 
strong to cancel it. Note that we have assumed (close to) 
symmetric $m_D$, as implied for instance by 
a discrete LR symmetry. If the symmetry basis is not the basis in which the 
charged lepton mass matrix $m_\ell$ is real and diagonal, 
then $m_D$ will be slightly non-symmetric if the matrix 
diagonalizing $m_\ell$ contains only 
small mixing angles. Our arguments would remain valid in this case. 
We should remark here that for highly 
non-symmetric Dirac mass matrices it is possible 
to generate successful large neutrino mixing even if $M_R$ and 
$m_D$ have very similar hierarchy \cite{ibarra}. 
To generate maximal mixing from Eq.~(\ref{eq:mnu22}) 
one would require $a = 1$, which means that two entries in $m_D$ 
are identical. The equality of certain elements is always necessary for 
extreme mixing angles. 
We can generalize the procedure to three generations. Suppose that 
$m_D$ is ``up-quark-like'', i.e., it 
contains masses (in units of $v$) of order 1, $\epsilon_D^2$ 
and $\epsilon_D^4$:  
\be
m_D = v \, 
\left( 
\bad
\epsilon_D^4 & a \, \epsilon_D^3 & b \, \epsilon_D^3 \\[0.2cm]
a \, \epsilon_D^3 & c \, \epsilon_D^2 & d \, \epsilon_D^2  \\[0.2cm]
b \, \epsilon_D^3 &  d \, \epsilon_D^2  & 1
\ea
\right) \mbox{ and } 
M_R = M \, 
\left( 
\bad
\epsilon_{M1} & 0 & 0 \\[0.2cm]
\cdot & \epsilon_{M2} & 0  \\[0.2cm]
\cdot & \cdot & 1 
\ea
\right)~,
\ee
where the diagonal $M_R$ is described by two small parameters 
$\epsilon_{M1}$ and $\epsilon_{M2}$. The light neutrino mass matrix is 
{\small 
\be
m_\nu = -\frac{v^2}{M} \, 
\left( 
\bad 
\epsilon_D^2 \left( b^2 \, \epsilon_D^4 + \eta_1 + a^2 \, \eta_2 \right) 
& \epsilon_D \left( b \, d  \, \epsilon_D^4 + a \, \eta_1 + a \, c \, 
\eta_2 \right) & 
 \epsilon_D \left( b \, \epsilon_D^2 + b \, \eta_1 + a \, d \, 
\eta_2 \right) \\[0.2cm]
\cdot & d^2 \, \epsilon_D^4 + a^2 \, \eta_1 + c^2 \, \eta_2 & 
d \, \epsilon_D^2 + a \, b \, \eta_1 + d \, c \, \eta_2 \\[0.2cm]
\cdot & \cdot & 1 + b^2 \, \eta_1 + d^2 \, \eta_2 
\ea 
\right)~,
\ee }
where we defined $\eta_1 = \epsilon_D^6 /\epsilon_{M1}$ and 
$\eta_2 = \epsilon_D^4 /\epsilon_{M2}$. To have a dominating 
23 block in this matrix, we can either have 
$\eta_2 \gg \eta_1, \epsilon_D^2$ or 
$\eta_1 \gg \eta_2, \epsilon_D^2$. In the first case we have 
\be
m_\nu = -\frac{v^2}{M} \, 
\left( 
\bad 
a^2 \, \epsilon_D^2 \, \eta_2 
& a \, c \, \epsilon_D \, \eta_2  & 
 a \, d \, \epsilon_D \, \eta_2  \\[0.2cm]
\cdot & c^2 \, \eta_2 & 
d \, c \, \eta_2 \\[0.2cm]
\cdot & \cdot & 1 + d^2 \, \eta_2 
\ea 
\right)~,
\ee
which for $\eta_2$ of order (or 
larger than one) is the wanted leading order structure 
of $m_\nu$. 
Note that maximal 23 mixing in case of $\eta_2 $ larger than one 
would require $d=c$, i.e., equality of certain mass matrix 
elements. 
Realistic predictions require corrections to this matrix 
from the remaining Majorana masses via $\eta_1$ \cite{sedom}. 
We stress again that this see-saw enhancement of the mixing 
requires that the hierarchy in $M_R$ is stronger, 
or the mixing is smaller, than that in the (close to symmetric) 
$m_D$. Consequently, 
if $M_R$ and $m_D$ have a similar flavor structure, and hence 
similar small mixing angles, then such a procedure is doomed. 
As we will argue in the following, in the type II see-saw 
case there is no problem in this case.

\subsection{The Situation in the Type II See-Saw}

We will show now that the peculiar interplay of the two 
terms in the type II see-saw formula can give large mixing even if 
$M_R$ and $m_D$ have small mixing of the same order of magnitude 
(``type II enhancement''). 
The need to construct models in which the flavor structure of the 
right-handed neutrinos is very much different from the one of the 
other fermions is therefore absent. 
What we essentially note is that if both $m_\nu^{II}$ and 
$m_\nu^I$ generate small mixing (as in the quark sector),  
their sum does not necessarily need to do so and can correspond to 
large neutrino mixing. 
The conditions under which this can occur are outlined below, but the 
essential requirement in our example is only a moderate 
cancellation in the 33 entry 
of $m_\nu$. 

Let us demonstrate the idea in a simple 2-neutrino 
framework: consider a hierarchical 
Dirac mass matrix of the form 
\be
m_D = v 
\left(
\bad
a_D \, \lambda^4  & b_D \, \lambda \\[0.2cm]
b_D \, \lambda    & 1  
\ea
\right)~.   
\ee
For simplicity, we have chosen here $m_D$ to be symmetric, an assumption 
which by no means affects the validity of our argument. 
Since the mechanism is working for similar flavor structures of the involved 
matrices, we do not introduce small $\epsilon_D$ for $m_D$ 
and $\epsilon_M$ for $M_R$, but rather parametrize the matrices in terms 
of a single small parameter $\lambda$, which can be 
thought of to be of the order of the Cabibbo angle. 
We choose a Majorana mass term for the right-handed neutrinos 
with similar hierarchy\footnote{If $m_D$ and $M_R$ had 
identical flavor structure, 
our argument would still work but  
the resulting formulae would become longer. 
We comment below on an interesting aspect of these scenarios.}: 
\be
M_R = v_R 
\left( 
\bad
a_R \, \lambda^3  & b_R \, \lambda \\[0.2cm]
\cdot & 1   
\ea
\right)~. 
\ee
We introduced real parameters $a_c$ and $b_c$ (with $c=D,R$ for the 
Dirac and Majorana mass matrix, respectively), which 
are of order one.  
Within the conventional see-saw mechanism, we have 
(giving only the lowest powers of $\lambda$) 
\be
m_\nu^I = -m_D^T \, M_R^{-1} \, m_D \simeq 
\frac{\D v^2}{\D v_R} \frac{\D b_D (b_D - 2 b_R)}{\D b_R^2}  
\left(
\bad
\frac{ \D a_R \, b_D}{\D b_D - 2 b_R} \lambda^3  
& \frac{ \D -b_D \, b_R}{\D b_D - 2 b_R} \lambda \\[0.4cm]
\cdot   & 1    
\ea
\right)~, 
\ee
which generates small mixing, $\theta^I = {\cal O}(\lambda)$. 
The mixing angles 
for $m_D$ and $M_R$, respectively, are also of the same order:  
$\theta^D \simeq \theta^R = {\cal O}(\lambda)$. 
Let us assume a discrete LR symmetry. Then, with $m^{II} \propto M_R$ 
we also have that the mixing of the triplet term is small: 
$\theta^{II} = {\cal O}(\lambda)$. 
Now we add $m_\nu^I$ to $m_\nu^{II}$, which yields  
\be \label{eq:mnu2} 
{\small 
m_\nu = m_\nu^{II} + m_\nu^I 
\simeq v_L 
\left( 
\baz 
\left(a_R + \frac{\D a_R \, b_D^2}{\D b_R^2 \, \gamma }\right) \, \lambda^3 & 
\left(b_R - \frac{\D b_D^2}{\D b_R \, \gamma}\right) \, \lambda \\[0.4cm]
\cdot & 
1 + \frac{\D b_D \, (b_D - 2 b_R)}{\D b_R^2 \, \gamma} + 
\frac{\D a_R \, (b_D - b_R)^2}{\D b_R^4 \, \gamma} \, \lambda
\ea 
\right)~.
} 
\ee
In general, this matrix generates two eigenvalues of 
order $1$ and $\lambda^2$ and a small mixing angle of order $\lambda$. 
The crucial observation for type II enhancement 
is the following: suppose 
the term of order $1$ in the $22$ element of $m_\nu$ 
in Eq.\ (\ref{eq:mnu2}) cancels. 
In this case one has large mixing 
given by 
\be
\tan 2 \theta \simeq \frac{4 \, b_D \, b_R^3}
{a_R \, (b_R - b_D)}= {\cal O}(1)~.  
\ee
If in addition $b_D = b_R$ holds we can even generate maximal mixing. 
Note that this formula holds when the order one term in 
Eq.\ (\ref{eq:mnu2}) cancels exactly. In order to generate large mixing 
(i.e., $\tan 2 \theta$ of order 1) 
it suffices however that the cancellation of the two terms generates a 
term of order $\lambda$. If the 11 entry is very small, it is crucial 
that cancellations make the 22 entry of $m_\nu^{II} + m_\nu^I$ have 
the same order as the 12 entry. In order not to ask for too strong 
cancellation (and therefore fine-tuning) there should 
before cancellation be only one order 
of magnitude difference between the 12 and 22 entry. 
We have assumed discrete LR symmetry, forcing $m_L$ and $M_R$ to 
be proportional to each other. This is obviously not necessary to make 
the mechanism work.\\

In the realistic case of three generations  
we wish to obtain now the leading structure of the low 
energy mass matrix corresponding to a normal hierarchy. 
It is therefore necessary that, after the cancellation, the lower 
23 block of $m_\nu$ has elements of the same order 
of magnitude, but larger than the entries in the first row.
One may choose the following structures of the mass matrices:
\be \label{eq:mdMR}
m_D = v \, 
\left( 
\bad 
a_D \, \lambda^4 & b_D \, \lambda^3 & c_D \, \lambda^3 \\[0.2cm]
\cdot & d_D \, \lambda^2 & e_D \, \lambda^2 \\[0.2cm]
\cdot & \cdot &  f_D
\ea 
\right) \mbox{ and } 
M_R = v_R \, 
\left( 
\bad 
a_R \, \lambda^3 & b_R \, \lambda^2 & c_R \, \lambda^2 \\[0.2cm]
\cdot & d_R \, \lambda & e_R \, \lambda \\[0.2cm]
\cdot & \cdot &  f_R
\ea 
\right) ~.
\ee
We also choose discrete LR symmetry which means here $m_L = v_L\, M_R /v_R$. 
The mass spectrum of $m_D$ is ``up-quark-like'', i.e., it 
contains masses (in units of $v$) 
of order 1, $\lambda^2$ and $\lambda^4$, while the eigenvalues of 
$M_R$ ($m_L$) are in units of $v_R$ ($v_L$)  
of order 1, $\lambda$ and $\lambda^3$. The two mass spectra of $m_D$ and $M_R$ 
are therefore similar, and small mixing is predicted by both 
matrices. 
The structure of the mass matrix in the conventional see-saw mechanism is 
\be
m_\nu^I \simeq \frac{v^2}{v_R}
\left( 
\bad 
{\cal O}(\lambda^5) & {\cal O}(\lambda^4) & {\cal O}(\lambda^3) \\[0.2cm]
\cdot & {\cal O}(\lambda^3) & {\cal O}(\lambda^2) \\[0.2cm]
\cdot & \cdot &  \frac{f_D^2}{f_R} + {\cal O}(\lambda) 
\ea 
\right) ~,
\ee
which can not reproduce the neutrino data. Hence, if only the 
conventional type I see-saw term or only the triplet term 
$m_L$ would contribute, then small neutrino mixing not capable 
of explaining the data would result. 
However, the total neutrino mass matrix reads 
\bea \label{eq:mnu3}
m_\nu = m_\nu^{II} + m_\nu^I \simeq v_L 
\left( 
\bad  
a_R \, \lambda^3 & b_R \, \lambda^2 &  c_R \, \lambda^2 \\[0.2cm]
\cdot & d_R \, \lambda & e_R \, \lambda \\[0.2cm]
\cdot & \cdot & (f_R - \frac{f_D^2}{f_R \, \gamma}) + \tilde{f} \, \lambda 
\ea
\right)~, 
 \\ \\
\mbox{ with } \tilde{f} = \frac{\D \left( c_R^2\,d_R -
       2\,b_R \,c_R \, e_R +
       a_R \, e_R^2 \right) \,f_D^2}{\D 
     \left( b_R^2 - a_R \, d_R \right) \, f_R^2\,\gamma}
~.
\eea
Let us again assume that the order one term in the 
33 entry of Eq.~(\ref{eq:mnu3}) cancels completely  
(again, it suffices that cancellation occurs just down to order $\lambda$).   
The condition for exact cancellation is quite simple, 
namely $f_R^2 = f_D^2/\gamma$, which is in 
fact simpler than the corresponding condition from the 2-flavor case 
discussed above. 
Then the mass matrix takes a well-known texture 
\be \label{eq:texture}
m_\nu \simeq v_L \, \lambda 
\left( 
\bad 
a_R \, \lambda^2 & b_R \, \lambda &  c_R \, \lambda \\[0.2cm]
\cdot & d_R & e_R \\[0.2cm]
\cdot & \cdot & \tilde{f}  
\ea 
\right)~, 
\ee
where the leading 23 block with entries of equal magnitude is 
necessary for large atmospheric mixing. 
The phenomenological consequences of Eq.~(\ref{eq:texture}) 
are a normal mass hierarchy 
with $m_3^2 \gg m_{2,1}^2$. Thereby renormalization effects are rendered 
subleading \cite{RGE}, unless in the MSSM with very large $\tan \beta$
(see below).  Moreover,  
\be
|U_{e3}| \sim \lambda \sim \sqrt{\frac{\dms}{\dma}}~, 
\ee
where $\dms \simeq 8 \cdot 10^{-5}$ eV$^2$ 
($\dma \simeq 2 \cdot 10^{-3}$ eV$^2$) governs the oscillations of solar and 
long base-line reactor (atmospheric and long-baseline accelerator) neutrinos. 
This fixes the magnitude of $\lambda \simeq 0.2$ and 
of $v_L \simeq \sqrt{\dma}/\lambda \simeq 0.2$ eV. 
Neutrinoless double beta decay is suppressed and triggered 
by a small effective mass of order 
$|m_{ee}| \sim \sqrt{\dma} \, |U_{e3}| \sim \sqrt{\dms}$.
The sizable $|U_{e3}|$ of order $\lambda$, 
a value close to current limits, is  
easily measurable in upcoming long-baseline or reactor 
oscillation experiments \cite{fut_meas}. 
Moreover, atmospheric neutrino mixing deviates sizably from maximal, 
$\tan 2 \theta_{23} \simeq 2 \, e_R/(\tilde{f} - d_R)$. This will be 
testable with future precision data, too. 
Without additional symmetries forcing some elements of $m_\nu$ 
to be equal, neither zero $|U_{e3}|$ nor maximal $\theta_{23}$ can be 
achieved in this framework. To be precise, in Eq.~(\ref{eq:texture}) 
one would need $b_R = c_R$ and $d_R = \tilde{f}$. 
Other aspects of type II see-saw, to be 
discussed in the next Sections, 
could be used to achieve extreme values of mixing angles.  

As is well known \cite{le}, the sub-determinant of the lower right 23 
block has to be of order $\lambda$ to generate a large solar neutrino mixing 
angle $\theta_{12}$. 
Therefore, two mild cancellations to order $\lambda$ are 
required: 
(i) the leading term in the 33 entry of $m_\nu$ 
has to cancel to order $\lambda$; 
(ii) the lower right 23 sub-determinant of $m_\nu$ has to be of order 
$\lambda$ to generate large solar neutrino mixing. 
The fact that two cancellations are required to make $\theta_{12}$ 
large, but only one to make $\theta_{23}$ large, 
could be used as an explanation why atmospheric neutrino mixing 
is larger than solar.\\

One may wonder whether one can generate the 
inverted hierarchy along similar lines. Here the requirement is 
that the 12 and 13 entries of $m_\nu$ are much larger than the other ones. 
When all individual matrices $m_L$, $M_R$ and $m_D$ correspond to 
small mixing, this would be rather unnatural since it requires cancellation 
within several independent elements of the resulting $m_\nu$. 
Similar statements can be made for quasi-degenerate neutrinos.
Note that we have chosen $m_L$ in a way that 
before cancellation there is only one order of magnitude 
difference between the 33 and the 22,23 elements of $m_\nu$. 
This guarantees that cancellation is necessary only for one 
entry. A more extreme example for cancellation in several elements 
can be found in Ref.~\cite{Xing}: a discrete LR symmetric 
type II see-saw model based on $S(3)_L \times S(3)_R$ 
was considered, which allows two terms for each Majorana mass matrix, 
one term proportional to the unit matrix and one proportional 
to the democratic 
matrix. The latter term appears in $m_\nu^I$ and $m_\nu^{II}$ and 
has to cancel in order to generate large neutrino mixing.\\

The condition for the 33 entry in $m_\nu$ of Eq.~(\ref{eq:mnu3}) to 
cancel down to order $\lambda$ can be written as 
$f_R - \frac{f_D^2}{f_R \, \gamma} 
\stackrel{!}{=} a \, \lambda$. One may wonder 
whether radiative corrections can lead to this condition. 
For the normal hierarchy and within the Standard Model the 
radiative effects are always negligible below 
the see-saw scale. This can change in case of 
the MSSM, however. The effect of radiative corrections below 
the see-saw scale is to multiply the 13 and 23 element with 
$(1 + \epsilon)$ and the 33 element with $(1 + \epsilon)^2$, where 
\[
\epsilon \simeq -(1 + \tan^2 \beta) \, \frac{m_\tau^2}{16 \, \pi^2 \, v^2} 
\, \ln \frac{M_X}{M_Z} \simeq -2 \cdot 10^{-5} \, (1 + \tan^2 \beta)~,
\]
with $m_\tau$ the tau lepton mass. Large values of $\tan \beta \gs 50$ 
could lead to a sizable correction $(1 + \epsilon) \sim \lambda$, but 
would cause this on all elements of the third column of $m_\nu$, 
in particular on the 23 entry. Hence, 
we cannot blame radiative effects for the generation of large mixing in 
the framework under study.\\

Our examples had slightly different textures for $m_D$ and 
$M_R \propto m_L$.  
However, one could imagine cases in which all matrices have 
identical powers of $\lambda$ in all entries. This implies an interesting 
aspect for the complete $6 \times 6$ neutrino mass matrix 
whose diagonalization will lead to Eq.\ (\ref{eq:II}): 
\be \label{eq:66}
{\cal M}_\nu =  \left( 
\baz 
m_L & m_D \\[0.3cm]
m_D^T & M_R 
\ea 
\right)~.
\ee
It is easy to show 
that if the structures of  
$m_L \propto M_R$ and $m_D$ are identical, i.e.\ $a_D = a_R$, $b_D = b_R$ 
and so on, then the determinant of ${\cal M}_\nu$ 
vanishes for $\gamma = 1$. The requirement for this is therefore 
that all matrices $m_L$, $M_R$ and $m_D$ are identical and only differ 
by their scales $v_L$, $v_R$ and $v$. 
In addition, the exact relation $v_L \, v_R = v^2$ must hold.

\section{\label{sec:II}Leading Structures for 
zero $U_{e3}$ and maximal $\theta_{23}$}

As mentioned above, with the type II see-saw enhancement discussed  
in the last Section, it is in general not possible to generate exactly 
maximal or zero mixing. 
Therefore, we will now assume that the leading 
structure of the neutrino mass matrix corresponds to 
zero $U_{e3}$ and maximal $\theta_{23}$ and is provided by one of the 
terms in the type II see-saw formula. 
Small corrections are supplied by the other term, which can be 
either subleading or of similar magnitude (a study with no corrections  
from the conventional see-saw term is given in \cite{aussies}). 
Let us recapitulate (see also \cite{andre} for the first 
two examples) the three simple, stable and 
often used candidates for zero $U_{e3}$ and maximal $\theta_{23}$: 
{\small \be \label{eq:cand}
({\rm A}): 
\sqrt{\frac{\dma}{4}} \, 
\left( 
\bad
0 & 0 & 0 \\[0.2cm]
\cdot & 1 & -1 \\[0.2cm]
\cdot & \cdot & 1 
\ea
\right)~,~
({\rm B}): 
\sqrt{\frac{\dma}{2}} \, 
\left( 
\bad
0 & 1 & 1 \\[0.2cm]
\cdot & 0 & 0 \\[0.2cm]
\cdot & \cdot & 0 
\ea
\right)~,~
({\rm C}): 
m_0 \, 
\left( 
\bad
1 & 0 & 0 \\[0.2cm]
\cdot & 0 & 1 \\[0.2cm]
\cdot & \cdot & 0 
\ea
\right)~.
\ee } 
They conserve the flavor charges $L_e$, $L_e - L_\mu - L_\tau$ and 
$L_\mu - L_\tau$, respectively. 
All three matrices have one eigenvalue with an eigenvector 
$(0,\,-1/\sqrt{2}, \,1/\sqrt{2})^T$. This eigenvalue is 
$\sqrt{\dma}$ for case (A), 0 for case (B) and $-m_0 $ for case (C). 
Therefore, they correspond 
to the normal hierarchy, the inverted hierarchy 
and quasi-degenerate neutrinos, respectively. 
Applying corrections to the three candidates is essential, since in 
their present form (A) and (B) have no solar $\Delta m^2$ while case (C) 
has no atmospheric $\Delta m^2$. Case (B) predicts maximal 
$\theta_{12}$, the other candidates have no physical 12 mixing. 
The matrices in Eq.~(\ref{eq:cand}) are exact, i.e., there are no order 
one coefficients involved. This is essential to have 
an eigenvalue of the form $(0,\,-1/\sqrt{2}, \,1/\sqrt{2})^T$ except for 
matrix (C). This is because $L_\mu - L_\tau$ is the only allowed $U(1)$ 
which is automatically $\mu$--$\tau$ symmetric \cite{mutau}.

One appealing possibility is that these simple matrices correspond to the 
triplet term $m_L$ and a small perturbation stems from the 
conventional see-saw term\footnote{A similar 
strategy to the one presented here 
has been discussed in the context of quasi-degenerate 
neutrinos previously in Ref.\ \cite{anki}. It was 
assumed that $m_L$ is proportional to the unit matrix (made possible 
by a $SO(3)$ symmetry) and the conventional see-saw term corresponds to 
sequential dominance \cite{sedom}, thereby generating quasi-degenerate 
neutrinos with large mixing.}. We thus assume that some symmetry 
enforces the triplet term to have one of the simple forms  
given in Eq.~(\ref{eq:cand}). The leading structures (A), (B) and (C) 
could also stem from the conventional see-saw term and the necessary 
correction from the triplet term. To generate such simple structures 
in $m_\nu^I$, interplay of the parameters in 
$m_D$ and $M_R$ is required. In a given theory or model this 
can be natural, but a priori it is more appealing that 
$m_L$ directly has this simple form. 

As already mentioned, we need to fill the zero entries in these 
matrices via the conventional see-saw term. 
In what regards the possibility of a discrete LR symmetry, it should 
be noted that cases (A) and (B) are singular and can not be inverted. 
Thus, if these matrices correspond to $m_L$, and if $M_R \propto m_L$, 
we can not construct the inverse of $M_R$ and the see-saw formula does 
not apply. Sterile neutrinos are the consequence of such a situation, 
for recent analyzes see \cite{sing}. One will have to omit the simplifying 
assumption $M_R \propto m_L$ in order to allow for a correction 
to the leading structure in $m_L$. On the other hand, the matrix (C) 
is invertible and can correspond to the triplet term in a 
discrete LR symmetric theory. Anyway, for simplicity and illustration 
we will focus here on three rather simple perturbations to the 
candidate matrices. What we mean by this is that $m_\nu^I$ 
has entries of the same order of magnitude, at most differing 
from each other by order one coefficients. The 
first possible perturbation is purely anarchical \cite{uk}: 
\be \label{eq:ana}
m_\nu^I \simeq  v_L \, \epsilon \,
\left( 
\bad
a & b & c \\[0.2cm]
\cdot & d & e \\[0.2cm]
\cdot & \cdot & f 
\ea
\right)~.
\ee 
Such a matrix can be obtained if both $m_D$ and $M_R$ are anarchical, 
or if only one of them is anarchical and the other one 
proportional to the unit matrix. 
The second perturbation corresponds to a $\mu$--$\tau$ 
symmetric \cite{mutau} matrix, i.e., 
$b = c$ and $d = f$: 
\be \label{eq:mt}
m_\nu^I \simeq  v_L \, \epsilon \,
\left( 
\bad
a & b & b \\[0.2cm]
\cdot & d & e \\[0.2cm]
\cdot & \cdot & d 
\ea
\right)~.
\ee 
As the candidate matrices in Eq.~(\ref{eq:cand}) are also 
$\mu$--$\tau$ symmetric, adding this perturbation will not change 
the values $U_{e3}=0$ and $\theta_{23} = \pi/4$. 
Such a $\mu$--$\tau$ symmetric correction can be 
achieved when $m_D$ and $M_R$ have a 23 exchange symmetry \cite{rabimt}. 
The third case occurs when all entries in 
$m_\nu^I$ are identical, i.e., $m_\nu^I$ is flavor democratic \cite{demo}:
\be \label{eq:demo}
m_\nu^I \simeq  v_L \, \epsilon \,
\left( 
\bad
1 & 1 & 1 \\[0.2cm]
\cdot & 1 & 1 \\[0.2cm]
\cdot & \cdot & 1 
\ea
\right)~.
\ee 
Symmetries such as $S(3)$ can lead to such a structure.\\

We start with the leading structure (A), corresponding to a 
normal hierarchy. If it would correspond to 
the triplet term $m_L$, one would have $v_L \simeq \sqrt{\dma}/2$. 
The perturbation has to generate the solar mass squared difference, 
therefore $\epsilon \simeq \sqrt{\dms/\dma}$. 
For an anarchical perturbation as in Eq.~(\ref{eq:ana}), one finds 
naturally large $\theta_{12}$, while 
$U_{e3} \simeq  \epsilon \, (b - c)/\sqrt{8}$, 
$\theta_{23} - \pi/4 \simeq \epsilon \, (d - f)/4 $ and 
$\dms/\dma \propto \epsilon^2$. 
Hence, both $U_{e3}$ and $\theta_{23} - \pi/4$ are of
order $\sqrt{\dms/\dma}$.
If $f = d$ one keeps  
$\theta_{23}$ maximal while $U_{e3} \neq 0$, and for $b = c$ it 
holds that $U_{e3}$ is zero while $\theta_{23} \neq \pi/4$ \cite{rabi}. 
Such a simple possibility does not exist for the other candidates 
(B) and (C). 
Both observables remain exactly zero if the type I correction 
is $\mu$--$\tau$ symmetric as in Eq.~(\ref{eq:mt}). 
Solar neutrino mixing is then naturally of order one: 
$\sin^2 \theta_{12} \simeq (a - d - e + w)/(2 \, w)$, where 
$w = \sqrt{8 \, b^2 + (a - d - e)^2}$. 
Now consider the flavor democratic perturbation 
from Eq.~(\ref{eq:demo}). One eigenvalues is zero, and one is 
$3 \, \epsilon \, v_L$ 
with an eigenvector $(1/\sqrt{3}, \, 1/\sqrt{3}, \,1/\sqrt{3})^T $ 
and therefore $\sin^2 \theta_{12} = \frac 13$. 
This is of course tri-bimaximal mixing \cite{tri}. 
We will elaborate more on this interesting possibility in 
Section \ref{sec:tbm}.\\ 

Let us turn to the inverted hierarchy. 
If matrix (B) corresponds to $m_L$, then $v_L \simeq \sqrt{\dma}/\sqrt{2}$. 
The correction to the zeroth order matrix (B) -- in absence of charged lepton 
contributions to the mixing matrix -- has to be sizable and tuned. 
It is however possible that an anarchical perturbation from $m_\nu^I$, 
being suppressed with respect to $m_L$ by a small factor of 
$\epsilon \sim \sqrt{\dms/\dma}$, 
corrects case (B) in an appropriate way, leading to 
$U_{e3}$ and $\theta_{23} - \pi/4$ of order $\sqrt{\dms/\dma}$. 
For a $\mu$--$\tau$ symmetric small correction one has that 
the smallest mass is $(d - e) \, \epsilon$ and of course 
$U_{e3} = \theta_{23} - \pi/4 = 0$. The ratio of mass squared 
differences is $\dms/\dma \simeq \sqrt{2} \, (a + d + e) \epsilon$ and 
solar neutrino mixing is governed by 
$\sin \theta_{12} \simeq \sqrt{\frac 12} - (a - d - e)\,\epsilon/8$. 
If the order one coefficients conspire such that 
$(a + d + e) \ll (a - d - e)$, then small \dms~goes along with non-maximal 
$\theta_{12}$. 
This in turn means that a flavor democratic perturbation 
does not work, since in this case 
$\dms/\dma \simeq 3 \sqrt{2} \, \epsilon$ and  
$|U_{e2}| \simeq \sqrt{\frac 12} - \epsilon/8$. Hence, 
$\sin \theta_{12} \simeq \sqrt{\frac 12} \, (1 - \frac{1}{24}\dms/\dma)$, 
which is too small a value. 

Apart from anarchical corrections, note that one needs a type I 
contribution of the form 
\be
m_\nu^I = -m_D^T \, M_R^{-1} \, m_D \simeq v_L 
\left( 
\bad
{\cal O}(\lambda) \mbox{ or } 
{\cal O}(1) & {\cal O}(\lambda^{n_1}) & {\cal O}(\lambda^{n_2})  \\[0.2cm]
\cdot &  {\cal O}(\lambda^{n_3}) & {\cal O}(\lambda^{n_4})  \\[0.2cm]
\cdot & \cdot & {\cal O}(\lambda^{n_5})
\ea
\right)~, 
\ee
where $n_i$ is some integer number. 
This implies non-trivial structures of $m_D$ and/or $M_R$. 
For instance, if (other choices are of course possible) 
\be
m_D = v \, 
\left( 
\bad 
a_D \, \lambda^4 & b_D \, \lambda^5 & c_D \, \lambda^5 \\[0.2cm]
\cdot & d_D \, \lambda^2 & e_D \, \lambda^2 \\[0.2cm]
\cdot & \cdot &  f_D \, \lambda 
\ea 
\right) \mbox{ and } 
M_R = v_R \, 
\left( 
\bad 
a_R \, \lambda^7 & 0 & 0 \\[0.2cm]
\cdot & d_R \, \lambda^2 & 0 \\[0.2cm]
\cdot & \cdot &  f_R
\ea 
\right) ~, 
\ee
we would get  
\be
m_\nu^I = -m_D^T \, M_R^{-1} \, m_D \simeq \frac{v^2}{v_R}
\left( 
\bad
{\cal O}(\lambda) & {\cal O}(\lambda^2) & {\cal O}(\lambda^2)  \\[0.2cm]
\cdot &  {\cal O}(\lambda^2) & {\cal O}(\lambda^3)  \\[0.2cm]
\cdot & \cdot & {\cal O}(\lambda^2)
\ea
\right)~,
\ee
which can satisfy the data if added to $m_L$.\\

Now, turning to quasi-degenerate neutrinos, 
assume that matrix (C) corresponds to $m_L$. An anarchical 
perturbation allows for successful phenomenology. 
Diagonalizing matrix (C) plus a flavor democratic perturbation, 
gives eigenvalues $-1$, 1 and $1 + 3 \epsilon$, where the latter 
has an eigenvector  $(1/\sqrt{3}, \, 1/\sqrt{3}, \,1/\sqrt{3})^T $, 
thereby resembling tri-bimaximal mixing. Recall that 
to accommodate the data, it is necessary that the 
neutrino with mass $m_2$ has this eigenvector. Thus, additional 
breaking is required (for instance via radiative corrections), 
in addition also because only one non-zero $\Delta m^2$ is present. 

Another possibility is the following: 
since the matrix (C) is invertible, we can assume discrete LR symmetry and 
thus $m_L \propto M_R$. Choosing for instance 
\be
m_D = v \, 
\left( 
\bad
a_D \, \lambda^3 & b_D \, \lambda^2 & c_D \, \lambda^2 \\[0.2cm]
\cdot & d_D \, \lambda & e_D \, \lambda \\[0.2cm]
\cdot & \cdot & f_D 
\ea
\right) \mbox{ and } 
M_R = v_R \, 
\left( 
\bad
X & 0 & 0 \\[0.2cm]
\cdot & 0 & Y \\[0.2cm]
\cdot & \cdot & 0 
\ea
\right)~,
\ee
gives a low energy mass matrix capable of explaining the 
data \cite{lmlt0,lmlt}: 
\be \label{eq:30}
m_\nu^{II} + m_\nu^{I} \simeq v_L \, 
\left( 
\bad
X & {\cal O}(\lambda^3) & {\cal O}(\lambda^2) \\[0.2cm]
\cdot & {\cal O}(\lambda^2) & Y \\[0.2cm]
\cdot & \cdot & {\cal O}(\lambda)
\ea
\right)~,
\ee
where only the leading terms are given. 
The Dirac mass matrix resembles the up-quarks and 
the form of $M_R$ is trivial to obtain if the heavy neutrinos 
$N_1$, $N_2$ and $N_3$ have the charges 0, 1 and $-1$ under 
$L_\mu - L_\tau$ \cite{lmlt0}.\\

We were using in this Section, in particular for the normal hierarchy, 
mainly a more or less anarchical 
perturbation generated by the type I see-saw term, which is 
somewhat incompatible with the naive expectation of hierarchical 
Dirac mass matrices and also with a discrete LR symmetry. 
In the next Section we will show that it is also possible 
to perturb a given mixing scenario when both hierarchical 
Dirac mass matrices and discrete LR symmetry are present. 
In this case the zeroth order mass matrix 
as provided by $m_L$ has to have a more complicated form.

\section{\label{sec:III}Deviations from Bimaximal and Tri-bimaximal Mixing}
In the last Section we have perturbed very simple mass 
matrices leading to $U_{e3}=0$ and $\theta_{23} = \pi/4$ via 
more or less anarchical perturbations from $m_\nu^I$. In particular, $m_D$ was 
required to possess a rather unusual structure. 
We show in this Section an alternative possibility to deviate 
(in the normal hierarchy) within the type II see-saw mechanism 
certain neutrino mixing scenarios, such as bimaximal \cite{bima} 
or tri-bimaximal \cite{tri} mixing. 
The difference with respect to the proposals in Section \ref{sec:II} 
is that the zeroth order mass matrix, as provided by $m_L$ 
has a more complicated structure. 
One can again imagine that these simple scenarios are implemented by some 
symmetry only in $m_L$, whereas the other mass matrices are connected 
to the ``hierarchical with small mixing'' form known from the quarks. 
In contrast to Section \ref{sec:II}, the perturbation 
generated by the conventional see-saw term works with a discrete 
LR symmetry and also with a hierarchical Dirac mass matrix \cite{ich2}. 
Both the bimaximal and tri-bimaximal scenario predict 
vanishing $\theta_{13}$ and $\cos 2 \theta_{23}$, therefore they are special 
cases of $\mu$--$\tau$ symmetry \cite{mutau,rabimt,lmlt0}. 
In general, the procedure 
described here will be possible for any $\mu$--$\tau$ symmetric mixing 
scenario, but for definiteness we stick to bimaximal and 
tri-bimaximal mixing. The latter 
is in perfect agreement with current data and 
a perturbation due to type II see-saw (or some other mechanism) 
is strictly speaking not necessary, 
but will lead to non-vanishing 
$\theta_{13}$ and $\cos 2 \theta_{23}$. In contrast to this, 
bimaximal mixing is ruled out by several standard deviations, and 
therefore requires a perturbation. As we will show, this 
perturbation can mimic Quark-Lepton Complementarity.\\

Let us start with bimaximal mixing \cite{bima}, defined as 
\be
\label{eq:Ubimax}
U_{\rm bimax} = 
\left(
\bad  
\frac{1}{\sqrt{2}} &  \frac{1}{\sqrt{2}} &  0 \\[0.3cm]
-\frac{1}{2} &  \frac{1}{2} &  -\frac{1}{\sqrt{2}} \\[0.3cm]
-\frac{1}{2} &  \frac{1}{2} &  \frac{1}{\sqrt{2}} \\[0.3cm]
\ea 
\right)~,
\ee
corresponding to 
$\theta_{12} = \theta_{23}=\pi/4$, $U_{e3}=0$ and  
leading to a mass matrix 
\bea \label{eq:mnubimax}
m_\nu^{\rm bimax} = \left( \bad 
\D A & B & B \\[0.2cm] 
\D \cdot &  \frac 12 \, (A + D)  &  \frac 12 \, (A - D)\\[0.3cm]
\D \cdot & \cdot &   \frac 12 \, (A + D)
\ea   \right)~, 
\eea
where 
\be \label{eq:ABD}
A = \frac{m_1^0 + m_2^0 }{2}~,~
B = \frac{m_2^0 - m_1^0}{2 \, \sqrt{2}}~
,~D = m_3^0~. 
\ee
The superscript 0 indicates that these are the initial 
mass eigenvalues, valid before a perturbation from the type II see-saw term 
is switched on. 
We demonstrate now how the type II see-saw mechanism can lead to 
a deviation from bimaximal mixing in accordance with neutrino data. 
We can assume again discrete LR symmetry, Eq.~(\ref{eq:mnuLR}), where 
$v_L \, f$ is now given by Eq.~(\ref{eq:mnubimax}). 
The inverse of 
$M_R$ is given by 
\bea 
M_R^{-1} = \frac{\D v_L}{\D v_R}~m_L^{-1} = 
\frac{\D v_L}{\D v_R}~
\left( \bad 
\tilde{A} & \tilde{B} & \tilde{B}  \\[0.3cm]
\cdot & \frac 12 (\tilde{A} + \tilde{D})
&   \frac 12 (\tilde{A} - \tilde{D})\\[0.3cm] 
\cdot & \cdot &   \frac 12 (\tilde{A} + \tilde{D}) \\[0.3cm] 
\ea \right)~,
\eea 
where 
\be
\tilde{A} = \frac{A}{A^2 - 2 B^2}~,~ 
\tilde{B} = \frac{-B}{A^2 - 2 B^2}~,~
\tilde{D} = \frac{1}{ D}~. 
\ee
We shall assume in the following a normal hierarchical mass spectrum, i.e., 
$(m_3^0)^2 \gg (m_{1,2}^0)^2$. For zero $m_1^0$ the mass 
matrix Eq.~(\ref{eq:mnubimax}) would be singular. 
Assuming that $m_D$ is hierarchical can be 
quantified as $m_D \simeq {\rm diag}(0,0,m)$.  It is then easy to show 
that the effect of the 
conventional see-saw term is only \cite{IIreco0,ich2}:
\be \label{eq:s} 
m_D^T \, M_R^{-1} \, m_D \simeq 
\left( 
\bad
0 & 0 & 0 \\[0.2cm]
\cdot & 0 & 0 \\[0.2cm]
\cdot & 0 & s 
\ea 
\right) ,\mbox{ where } 
s \equiv v_L^2~\frac{m^2}{4 \gamma \, v^2} 
\left(\frac{1}{m_1^0} + \frac{1}{m_2^0} + 
\frac{2 }{m_3^0} \right)~.
\ee
This term has to be subtracted from $m_L$ which is 
given in Eq.~(\ref{eq:mnubimax}). The zero entries in this matrix can 
also be small and suppressed with respect to the 33 element without 
changing our conclusions. 
With $\gamma \simeq 1$, $m \simeq v$ and one of the 
$m_i^0$ of order $v_L$, this conventional term is 
of similar magnitude as the triplet contribution, which is 
proportional to $v_L$. Hence, identifying $m_D$ 
with the charged leptons or the down-quarks will lead to a negligible 
correction of $m_\nu^I$ to $m_\nu^{II}$ if $\gamma = {\cal O}(1)$. 
If however $m_D$ is related to the up-quarks, 
$m \simeq v$, then we can estimate this term as 
\be \label{eq:sapp}
s \simeq \frac{0.1}{4 \, \gamma} \left( \frac{v_L}{10^{-2}~ \rm eV} \right)^2 
\left( \frac{10^{-3}~ \rm eV}{m_1^0} \right) ~{\rm eV}~, 
\ee 
where again hierarchical $m_i^0$ were assumed. Leaving LR symmetry aside, 
many non-singular mass matrices $M_R$ in connection with hierarchical 
Dirac mass matrices will have the 33 entry of $m_\nu^I$ 
as the leading term and can be cast in the form (\ref{eq:s}). 
Naturally, for reference values $m_1^0 = 10^{-3}$ eV and $v_L = 10^{-2}$ eV,  
the order of $s$ can be -- without varying $\gamma$ around the 
value one within more than 
one order of magnitude --  
given by the scale of 
neutrino masses $\sqrt{\dms}$ or $\sqrt{\dma}$. 

We can now diagonalize the perturbed mass matrix. 
For $s$ of order $D$ or smaller and for $D^2 \gg A^2, B^2$ 
the mixing angles are given by 
\be \label{eq:CPCmix} \D 
|U_{e3}| \simeq \frac{B \, s}{\sqrt{2} \, D^2}~,~
\sin^2 \theta_{23} \simeq \frac 12 \, \left(1 + \frac{s}{D} \right) ~,~
\tan 2 \theta_{12} \simeq 4\sqrt{2} \, \frac{\D B}{\D s}~.
\ee  
From the expression for $\theta_{12}$ and assuming hierarchical 
$m_{i}^0$, one obtains that $|s| \sim |m_2^0| \sim \sqrt{\dms}$ in order 
to reproduce the observations. One interesting aspect, which we 
will assume now, is the following: 
from Eqs.\ (\ref{eq:mnubimax}) and (\ref{eq:ABD}) it is obvious 
that for $m_1^0 = - m_2^0$, or $A = 0$, one would start with vanishing $\dms$. 
In this case the conventional see-saw term $s$ 
creates not only the required deviation from maximal solar neutrino 
mixing, but induces also the solar mass squared difference, which is 
then proportional to $s^2$. The phenomenological relation that 
the deviation from maximal solar neutrino mixing is of the same order as  
$\sqrt{\dms/\dma}$ can thereby be explained, since the same parameter 
is responsible for both deviations.\\

We can discuss also a possible connection to 
Quark-Lepton Complementarity (QLC) \cite{QLC}. 
The deviation from maximal solar neutrino mixing can empirically 
be written as \cite{WR} $U_{e2} = \sqrt{1/2}~(1 - \lambda)$, where 
$\lambda \simeq 0.22$ quantifies the required deviation. 
If not a coincidence, the parameter $\lambda$ is the 
sine of the Cabibbo angle $\theta_{C}$ and therefore \cite{QLC} 
$\theta_{12} + \theta_{C} = \pi/4$. 
In this case $\tan 2 \theta_{12} = 1/(2 \lambda) + {\cal O}(\lambda)$, 
and from comparing Eq.~(\ref{eq:CPCmix}) with this expression it 
follows that QLC is mimicked\footnote{Strictly speaking, every 
model predicting $\sin^2 \theta_{12} \simeq 0.28$ mimics QLC, in the sense 
that this is about the prediction of $\theta_{12} = \pi/4 - \theta_C$. 
What we mean here by mimicking QLC is that one gets from bimaximal 
mixing to $\sin^2 \theta_{12} \simeq 0.3$.} 
when $s/B \simeq 8 \sqrt{2} \, \lambda$. 
In order to distinguish the type I contribution to bimaximal mixing 
from QLC, we note that 
there are two main scenarios in which QLC can arise 
\cite{QLC} (a recent detailed analysis of the low and high energy 
phenomenology of these two scenarios has been conducted in \cite{hr}). 
Their most important and most easily testable difference is that 
one scenario predicts $|U_{e3}|^2 = \lambda^2/2  \simeq 0.03$, 
while the other one 
predicts $|U_{e3}| = A \, \lambda^2 /\sqrt{2} \simeq 0.03$, 
where $A$ is a parameter 
in the Wolfenstein parametrization of the CKM matrix. In our framework, 
one finds that $|U_{e3}|$ is of similar size than in the second QLC 
scenario, but obeys the correlation 
\be
\frac{2 \, |U_{e3}|}{\tan 2 \theta_{12}} \simeq 
\left( \sin^2 \theta_{23} - \frac 12 \right)^2 \simeq 
\frac{\dms}{\dma} \, \cos 2 \theta_{12} ~. 
\ee
Since this relation is not predicted by the QLC scenario, 
we can in principle distinguish it from our scenario.

Neutrino mixing can also be very well described by tri-bimaximal mixing 
\cite{tri}, which is defined by the mixing matrix in Eq.~(\ref{eq:Utri}). 
The resulting mass matrix $m_\nu = U^\ast \, m_\nu^{\rm diag} \, U^\dagger$ 
is 
\be \label{eq:mnuTBM}
m_\nu = 
\left(
\bad 
A & B & B \\[0.2cm]
\cdot & \frac{1}{2} (A + B + D) & \frac{1}{2} (A + B - D)\\[0.2cm]
\cdot & \cdot & \frac{1}{2} (A + B + D)
\ea 
\right)~, 
\ee
where 
\be \label{eq:ABD1}
A = \frac{1}{3} (2 \, m_1^0 +  m_2^0)~,~~
B = \frac{1}{3} (m_2^0 - m_1^0)~,~~
D =  m_3^0~, 
\ee
or $m_1^0 = A - B$ and $m_2^0 = A + 2 \, B$. For normal 
hierarchical neutrinos we have $D^2 \gg A^2, B^2$. 
Note that if we remove $B$ from the 
23 block of $m_\nu$ we obtain Eq.~(\ref{eq:mnubimax}), i.e., 
bimaximal mixing. 
Suppose again that the mass matrix (\ref{eq:mnuTBM}) 
corresponds to $m_\nu^{II}$. In analogy to the example for bimaximal 
mixing given above the conventional see-saw term will result in 
a small contribution to 
the 33 entry. 
The results for $U_{e3}$ and $\theta_{23} - \pi/4$ are similar to the 
case of initial bimaximal mixing discussed above, while for solar 
neutrino mixing it holds 
\be \label{eq:NHresmnu_mt}
\tan 2 \theta_{12} \simeq \frac{2\sqrt{2}}{1 - s/(2 B)} ~.
\ee 
A slightly smaller $s$ is required in this case, which can be 
expected, since tri-bimaximal mixing is very close to current data and little 
room for deviations is there.

\section{\label{sec:tbm}More on Tri-bimaximal Mixing} 
We have seen in Section \ref{sec:II} that a sum of two 
relatively simple matrices can lead to (close to) tri-bimaximal mixing in 
the normal hierarchy. 
We will now comment more on the realizations of 
this mixing scheme within the type II see-saw, discussing also the 
inverted hierarchy and quasi-degenerate neutrinos. 
Tri-bimaximal mixing is defined as \cite{tri} 
\be \label{eq:Utri} 
U = \left(
\bad 
\sqrt{\frac{2}{3}} & \sqrt{\frac{1}{3}} & 0 \\[0.2cm]
-\sqrt{\frac{1}{6}} & \sqrt{\frac{1}{3}} & -\sqrt{\frac{1}{2}}  \\[0.2cm]
-\sqrt{\frac{1}{6}} & \sqrt{\frac{1}{3}} & \sqrt{\frac{1}{2}}  
\ea 
\right)~. 
\ee
It corresponds to 
$\sin^2 \theta_{12} = 1/3$, $U_{e3}=0$ and $\theta_{23}=\pi/4$. The 
resulting mass matrix $m_\nu = U^\ast \, m_\nu^{\rm diag} \, U^\dagger$ 
can be written in terms of matrices multiplied with the masses: 
\be
m_\nu = \frac{m_1}{6} \, 
\left( 
\bad 
4 & -2 & -2 \\
\cdot & 1 & 1 \\
\cdot & \cdot & 1 
\ea 
\right) 
+ \frac{m_2 }{3} \, 
\left( 
\bad 
1 & 1 & 1 \\
\cdot & 1 & 1 \\
\cdot & \cdot & 1 
\ea 
\right) + 
\frac{m_3 }{2} \, 
\left( 
\bad 
0 & 0 & 0 \\
\cdot & 1 & -1 \\
\cdot & \cdot & 1 
\ea 
\right)~.
\ee
This equation is exact, i.e., there are no order one coefficients involved. 
We remark here that such a sum of three matrices could also be realized 
if there are three different contributions to the effective mass 
matrix, such as the ones mentioned at the end of Section \ref{sec:IIgen}.
When we assume a normal hierarchy and neglect $m_1$ it follows 
\be
m_\nu = 
 \frac{\sqrt{\dms}}{3} \, 
\left( 
\bad 
1 & 1 & 1 \\
\cdot & 1 & 1 \\
\cdot & \cdot & 1 
\ea 
\right) + 
\frac{\sqrt{\dma}}{2} \, 
\left( 
\bad 
0 & 0 & 0 \\
\cdot & 1 & -1 \\
\cdot & \cdot & 1 
\ea 
\right)~,
\ee
i.e., the second term dominates. One of the terms 
could stem from the triplet term and the other one from the 
conventional see-saw term. In Section \ref{sec:II} we encountered 
this matrix when we perturbed a triplet term corresponding to 
the atmospheric neutrino mass scale (matrix (A) in Eq.~(\ref{eq:cand})) 
with a flavor democratic  
type I term. 
Another possibility is that 
the triplet term is subleading and corresponds to the 
flavor democratic contribution 
proportional to $\sqrt{\dms}$.  
The second, leading term is then generated by the conventional see-saw 
mechanism, for instance via sequential dominance or conservation 
of $L_e$.  Note however that a democratic mass matrix 
has rank 1 and can not be inverted. Hence, if there is a discrete 
LR symmetry then the term proportional to the democratic matrix can not stem 
from the triplet but must come from the conventional see-saw term. 
Further note that since the non-vanishing entries of the second term 
are identical, there will be additional symmetries, such as $S_2$ or $Z_2$, 
required. For instance, if 
the democratic term is generated by a triplet term, then the second 
term could stem from $M_R \propto {\rm diag}(0,0,1)$ and for $m_D$ 
it suffices that the third row looks like $(0,-1,1)$.\\

Another possibility is that there are similar contributions 
of the type I and triplet term. We can then discuss the inverted hierarchy 
and quasi-degenerate neutrinos. 
With $m_3 = 0$ and equal $CP$ parities of the remaining states, i.e., 
$m_1 = m_2$, we can write 
\be \label{eq:46}
\frac{m_\nu}{\sqrt{\dma}} \simeq  
\frac 12 
\left( 
\bad 
1 & 0 & 0 \\
\cdot & 1 & 0 \\
\cdot & \cdot & 1  
\ea 
\right) + 
\frac 12 
\left( 
\bad 
1 & 0 & 0 \\
\cdot & 0 & 1 \\
\cdot & \cdot & 0  
\ea 
\right)~.
\ee
Note that $m_\nu^I$ and $m_\nu^{II}$ have to have 
almost the same size.
Strictly speaking, $\theta_{12}=0$ results from this matrix. 
The underlying reason is the simplifying 
assumption $m_1 = m_2$, for which small corrections 
(of order $\dms/m_1^2$) are neglected. 
However, the quasi-degeneracy of the two neutrino masses can easily 
lead to large solar neutrino mixing once small breaking parameters are 
introduced. Breaking is necessary anyway in order to generate the 
solar mass splitting.  
Eq.~(\ref{eq:46}) is a sum of a unit matrix and a matrix conserving 
$L_\mu - L_\tau$ (plus an additional symmetry making the 11 and 
23 entries identical). 
The first matrix could be a triplet term, generated 
with $SO(3)$, and the 
second term could stem from a type I see-saw with 
a diagonal $m_D$ and a $M_R$ of the form 
\[
M_R = M \, 
\left( 
\bad 
1 & 0  & 0 \\
\cdot & 0 & 1 \\
\cdot & \cdot & 0  
\ea 
\right)~.
\] 
We encountered this kind of contribution from 
$m_\nu^I$ already at the end of Section \ref{sec:II}, see the remarks after 
Eq.~(\ref{eq:30}). 
It could also be that there is a discrete LR symmetry, in which 
$m_L$ and $M_R$ are proportional to the unit matrix. A very unusual 
form of $m_D$ is then required in order to obtain the second 
term in Eq.~(\ref{eq:46}). 
We could also generate this scenario when both 
$m_D$ and $M_R$ are proportional to the unit matrix 
and the triplet term obeys $L_\mu - L_\tau$. 
For an inverted hierarchy with opposite $CP$ parities, we have 
\be 
\frac{m_\nu}{\sqrt{\dma}} \simeq  
-\frac 23 
\left( 
\bad 
0 & 1 & 1 \\
\cdot & 0 & 0 \\
\cdot & \cdot & 0  
\ea 
\right) + 
\frac 13 
\left( 
\bad 
1 & 0 & 0 \\
\cdot & -\frac 12 &  -\frac 12 \\
\cdot & \cdot &  -\frac 12  
\ea 
\right)~, 
\ee
where both terms correspond to $L_e - L_\mu - L_\tau$. This 
matrix corresponds to tri-bimaximal mixing and requires small 
breaking in order to generate the solar mass splitting.\\ 

Similar discussions are possible for quasi-degenerate neutrinos. 
If $m_1 = m_3 = -m_2$, then we can write 
\be
\frac{m_\nu}{m_1} \simeq  
\frac 13  
\left( 
\bad 
1 & 0 & 0 \\
\cdot & 1 & 0 \\
\cdot & \cdot & 1  
\ea 
\right) -  
\frac 23 
\left( 
\bad 
0 & 1 & 1 \\
\cdot & 0 & 1 \\
\cdot & \cdot & 0  
\ea 
\right)
\mbox{ or } \frac{m_\nu}{m_1} \simeq  
\frac 13  
\left( 
\bad 
1 & 1 & 1 \\
\cdot & 1 & 1 \\
\cdot & \cdot & 1  
\ea 
\right) -  
\left( 
\bad 
0 & 1 & 1 \\
\cdot & 0 & 1 \\
\cdot & \cdot & 0  
\ea 
\right)~.
\ee
We can therefore express the mass matrix as a sum of a triangular matrix and 
a unit (or a democratic) matrix. 
For $CP$ parities leading 
to $m_1 = -m_3 = -m_2$ we can decompose the mass matrix as 
\be
\frac{m_\nu}{m_1} \simeq  
-\frac 23  
\left( 
\bad 
1 & 1 & 1 \\
\cdot & 1 & 1 \\
\cdot & \cdot & 1  
\ea 
\right) + 
 \left( 
\bad 
1 & 0 & 0 \\
\cdot & 0 & 1 \\
\cdot & \cdot & 0  
\ea 
\right)~,
\ee
where flavor democracy and $L_\mu - L_\tau$ seem to play a role again. 
Discrete LR symmetry is again possible.
The last two cases do not produce tri-bimaximal mixing, but can 
easily do so for appropriate small breaking parameters of order 
$\dms/m_1$ and $\dma/m_1$.

\section{\label{sec:concl}Summary}

Both the type I and the type II see-saw mechanism explain tiny neutrino 
masses, but large neutrino mixing is not predicted per se, unless there 
is additional input. While generating large neutrino mixing is well and 
often studied within the conventional (type I) see-saw mechanism, large 
or maximal mixing within the type II see-saw received so far little 
attention. Therefore we discussed in this article the interplay of both 
terms of the type II see-saw in order to understand the unexpected 
features of neutrino mixing. The fact that the neutrino mass matrix is 
in this case a sum of two terms opens up the possibility of cancellation 
if the two terms are comparable. It is also possible that the sum of 
two terms generates the unexpected features of neutrino mixing. Alternatively 
and most natural, one term can be the leading contribution, while the other 
one can give perturbations. 
In this context, several possibilities were suggested in this article: 
\begin{itemize}
\item we introduced ``type II enhancement'', i.e., showed that within 
type II see-saw models mild cancellation of certain terms can lead to 
the generation of large mixing angles, even though all individual matrices 
involved predict small mixing. Both discrete and gauge LR symmetry are 
possible. A hint to obtain such models is to note that the complete 
$6 \times 6$ neutrino mass matrix can have a vanishing determinant. 
The requirement that there is similar, but small, mixing in both
$m_D$ and $M_R$ differs from the usual (type I) see-saw enhancement 
of neutrino mixing, which requires a stronger hierarchy in the 
heavy neutrino sector and somewhat decouples the two sectors. 
Maximal or vanishing mixing requires additional input, such as the 
equality of certain mass matrix elements;   

\item the leading structure of the neutrino mass matrix as displayed 
in Eq.~(\ref{eq:cand}) can be generated by some symmetry acting on $m_L$. 
Necessary corrections stem from the conventional 
see-saw term. However, in case of a normal and inverted hierarchy 
the leading structures given in Eq.~(\ref{eq:cand}) 
(corresponding to $L_e$ and $L_e - L_\mu - L_\tau$, respectively) are 
singular, which make these scenarios incompatible with the discrete LR 
symmetric relation $m_L \propto M_R$. In contrast to this, the 
leading structure for quasi-degenerate neutrinos can be generated 
by the unit matrix or via 
the matrix in Eq.~(\ref{eq:lmlt}), corresponding to $L_\mu - L_\tau$. 
They can be inverted and are compatible with discrete LR symmetry.  
We showed that anarchical perturbations can generate successful 
phenomenology from the zeroth order matrices and that a 
$\mu$--$\tau$ symmetric perturbation keeps the initial values of 
zero $U_{e3}$ and maximal $\theta_{23}$; 

\item one could imagine that the triplet term 
has a more complicated structure corresponding to 
bimaximal or tri-bimaximal mixing. In discrete and gauge LR symmetric  
scenarios with hierarchical Dirac mass matrices it is easily possible that 
a small perturbation to $m_L$ arises, which deviates the mixing scenarios. 
Quark-Lepton Complementarity
could be mimicked, and in addition the empirical relation 
$1 - \sqrt{2} \, \sin \theta_{12} \simeq \sqrt{\dms/\dma}$ can 
be explained if $m_L$ alone would generate vanishing $\dms$; 

\item 
we realized that for tri-bimaximal mixing the light neutrino 
mass matrix can often be written as a sum of two terms both of 
which have an interesting structure.  
We interpret this by assuming that each term stems from one of 
the two terms in the type II see-saw formula. 
For a normal hierarchy, the two contributions have 
different order of magnitude, their ratio is given 
by $\sqrt{\dms/\dma}$. 
For the inverted hierarchy and for quasi-degenerate neutrinos, 
they have to have similar size. 

\end{itemize}

The next generation of experiments will show if $\theta_{13}$ 
is small or tiny and if $\theta_{23}$ is large or close to maximal. 
The discussed (incomplete) list of scenarios shows how the interference 
of the two terms in the type II see-saw leads to various interesting 
possibilities to understand all possibilities.

\begin{center}
{\bf Acknowledgments}
\end{center}
We thank R.~Mohapatra for discussions. 
This work was supported by the ``Deutsche Forschungsgemeinschaft'' in the 
``Transregio Sonderforschungsbereich TR 27: Neutrinos and Beyond'' 
and under project number RO--2516/3--2. 


\end{document}